# Beyond Social Fragmentation: Coexistence of Cultural Diversity and Structural Connectivity Is Possible with Social Constituent Diversity

Hiroki Sayama[1,2] and Junichi Yamanoi[1]

[1] Waseda Innovation Lab, Waseda University, Shinjuku, Tokyo 169-8050, Japan
[2] Center for Collective Dynamics of Complex Systems, Binghamton University, Binghamton NY 13902-6000, USA
sayama@binghamton.edu

**Abstract.** Social fragmentation caused by widening differences among constituents has recently become a highly relevant issue to our modern society. Theoretical models of social fragmentation using the adaptive network framework have been proposed and studied in earlier literature, which are known to either converge to a homogeneous, well-connected network or fragment into many disconnected subnetworks with distinct states. Here we introduced the diversities of behavioral attributes among social constituents and studied their effects on social network evolution. We investigated, using a networked agent-based simulation model, how the resulting network states and topologies would be affected when individual constituents' cultural tolerance, cultural state change rate, and edge weight change rate were systematically diversified. The results showed that the diversity of cultural tolerance had the most direct effect to keep the cultural diversity within the society high and simultaneously reduce the average shortest path length of the social network, which was not previously reported in the earlier literature. Diversities of other behavioral attributes also had effects on final states of the social network, with some nonlinear interactions. Our results suggest that having a broad distribution of cultural tolerance levels within society can help promote the coexistence of cultural diversity and structural connectivity.

## 1 Introduction

Social fragmentation caused by widening differences among constituents has recently become a highly relevant issue to our modern society, as various forms of gaps and conflicts are emerging from cultural, political, economic, ethnic, religious, linguistic, and other driving factors. Researchers have developed theoretical models of social fragmentation using the adaptive network framework [1-5], where the topologies of social ties between constituents and their states co-evolve simultaneously through homophily, social contagion, and/or other social processes. Such adaptive social net-work models are known to either converge to a homogeneous, well-connected network (= loss of cultural diversity), or fragment into many disconnected subnetworks with distinct states (= loss of structural connectivity). From a viewpoint of social capital and innovation, however, neither of these two social states would be desirable, because the former would mean the loss of information and the latter the loss of communication. In order to keep our society active and innovative, we should maintain cultural/informational diversity within our society high while also maintain information exchange and communication actively ongoing. This can be conceptualized as a structurally well-connected network with diverse node states. Earlier theoretical models of adaptive social network dynamics did not succeed in demonstrating how such outcomes could occur.

We note that those earlier models typically used stylized assumptions that behavioral attributes of social constituents were spatially homogeneous, and therefore, they may not have fully reproduced richer macroscopic outcomes, such as potential coexistence of diverse cultures within a connected network structure. To overcome this limitation, here we introduced the diversities of behavioral attributes among social constituents into an adaptive social network model, and computationally investigated their effects on social network evolution.

The rest of the paper is organized as follows. Section 2 describes our networked agent-based simulation model. Section 3 describes the design of our computational experiments and outcome measures. Section 4 summarizes the results. Section 5 concludes the paper with a brief discussion on the implications of the results for relevant research fields and real-world socio-cultural dynamics.

## 2 Model

For the purpose of this study, we developed a computational adaptive social network model of cultural diffusion dynamics by using our previous work on cultural integration in corporate merger [6,7] as the basis and implementing some revisions to it to allow representation of social constituent diversity.

In this model, we simulate the dynamics of an adaptive social network made of two initially distant cultural groups, each consisting of 50 individual constituents (nodes). Individuals are connected to each other through directed weighted edges, which represent the direction and intensity of cultural information flow. Constituents within each group and across the two groups are initially connected randomly with 20% and 2% edge densities, respectively, to represent initially modularized social structure. Edge weights are initially random, with weights sampled from a uniform distribution between 0 and 1. This initial network structure captures the state of two groups that is distant from each other both structurally and culturally.

Each individual constituent has its cultural state as a vector in a 10-dimensional continuous cultural space, based on previous empirical studies on measuring organizational cultural dimensions [8,9]. The distance between two cultures is characterized by the Euclidean distance between their two vectors in the cultural space. The cultural distributions among individual constituents are initialized as follows: First, two cul-tural "center" vectors are created for the two groups, separated by 3.0 (in an arbitrary unit) in the cultural space. Then individual cultural vectors are created for individuals in each group by adding a random number drawn from a normal distribution with a mean of 0 and a standard deviation of 0.1 (in the same unit used above) to each component of the cultural center vector of that group. This creates an initial condition in which the average between-group cultural difference is approximately seven times larger than the average within-group cultural difference.

Such an initial condition made of two distant, distinct cultural clusters may not be a popular choice for models studied in complex systems, network science, and statistical physics, where more randomized, homogeneous initial conditions are typically preferred. However, such random homogeneous conditions are extremely rare and unrealistic in real society, even for initial conditions. Rather, large-scale social systems emerge and evolve through numerous encounters and interactions between multiple smaller communities that are often culturally distinct from each other at the beginning. The heterogeneous, clustered initial conditions adopted in our study were intended to capture such social encounter situations, with the aim to increase the realism and applicability of our model and results in view of actual social self-organization and evolution.

Each iteration in simulation consists of simulating actions for all individual constituents in a sequential order. In its turn to take actions, an individual first selects an information source from its local in-neighbors with 99% probability (in this case the selection probabilities are proportional to edge weights), or with 1% probability, from anyone in the connected component to which the focal individual belongs. If there is no edge in the latter case, a new directed edge is created from the source to the focal individual with a minimal edge weight 0.01. Then, the individual decides to either accept or reject the source's cultural vector based on the distance between the received cultural vector and its own. The probability of cultural acceptance $P_A$ is an exponentially decreasing function of the cultural distance, defined as

$$P_A = \left(\frac{1}{2}\right)^{\frac{|v_i - v_j|}{d}} , \tag{1}$$

where $v_i$ and $v_j$ are the cultural vectors of the individual's own and of the selected source, respectively, and $d$ is the cultural tolerance, or the characteristic cultural distance at which $P_A$ becomes 50%. If the received culture is accepted, the individual's cultural vector is updated as

$$v_i \to (1 - r_s)v_i + r_s v_j , \tag{2}$$

where $r_s$ is the rate of cultural state change, and the edge weight from the source to the focal individual $w_{ij}$ is updated as

$$w_{ij} \to \text{logistic}\big(\text{logit}(w_{ij}) + r_w\big) , \tag{3}$$

where $r_w$ is the rate of edge weight change. Or, if the received culture is rejected, no change occurs to the focal individual's culture, but the edge weight is updated in an opposite direction as

$$w_{ij} \rightarrow \text{logistic}\left(\text{logit}\left(w_{ij}\right) - r_w\right) . \tag{4}$$

The above formula that combines logit and logistic functions guarantees that the updated edge weight is always constrained between 0 and 1. When the edge weight falls below 0.01, the edge is considered insignificant and is removed from the network. Additional details of these model assumptions, parameter settings, and their rationale can be found in our earlier work [6,7].

In the present study, we use $d = 0.5$, $r_s = 0.5$ and $r_w = 0.5$ as their mean values within the social network, and we systematically vary their variances among social constituents as the key experimental parameters. More details are given in the following section.

## 3 Experiments

We computationally investigated how the resulting social network states and topologies would change as social constituents' behavioral attributes were systematically diversified within the simulated society. The standard deviations of $d$, $r_s$, and $r_w$ were varied from 0 to 0.5 at interval 0.1, which makes the total number of parameter value combinations $|\{0, 0.1, 0.2, 0.3, 0.4, 0.5\}|^3 = 6^3 = 216$. We ran 100 independent simulation runs for each specific combination of parameter values (and therefore, the total number of simulations = 21,600 runs). Each run was simulated for 500 iterations.

After each simulation run was completed, the following two quantities were measured as outcome variables on the final network configuration:

- <*CD*> : average cultural distance between constituents in the initially distant two groups
- <*SPL*> : average shortest path length within the whole network

In these two measures, social fragmentation transitions can be captured as a positively correlated increase or decrease of both quantities. Namely, (<*CD*>, <*SPL*>) = (high, high) implies social fragmentation, while (<*CD*>, <*SPL*>) = (low, low) implies social assimilation with loss of cultural diversity.

## 4 Results

Figure 1 shows the baseline behaviors of the proposed model in the low behavioral diversity parameter region ($d \leq 0.1$, $r_s \leq 0.1$, $r_w \leq 0.1$), in which previously reported social fragmentation transitions are clearly observed as transitions between (<*CD*>, <*SPL*>) = (high, high) and (<*CD*>, <*SPL*>) = (low, low) behaviors (Fig. 1, black trend curve). Meanwhile, none of the simulation results showed (<*CD*>, <*SPL*>) = (high, low) behaviors (Fig. 1, red dashed circle) when social constituents were behaviorally homogeneous.

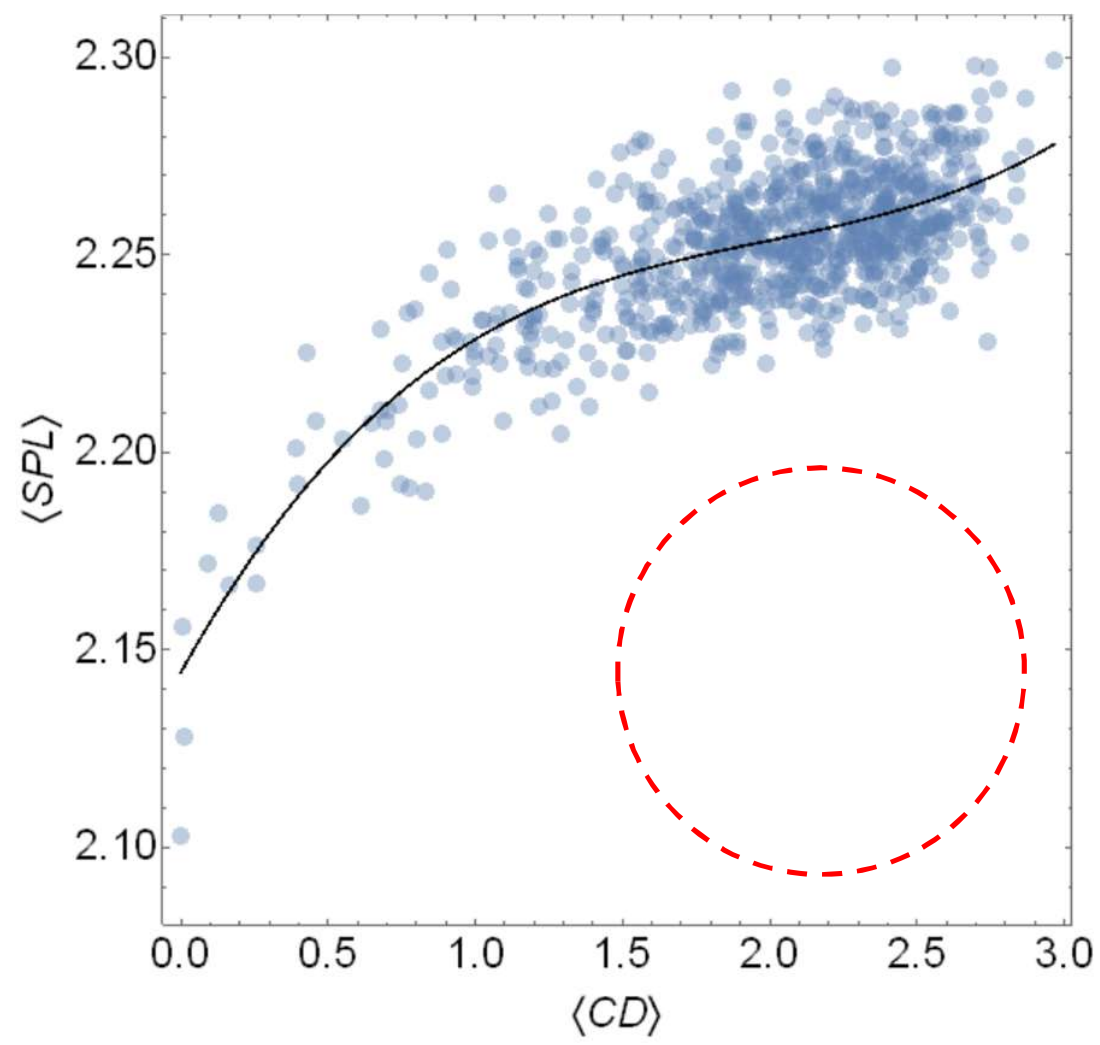

**Fig. 1.** Scatter plot showing two outcome measures <*CD*> and <*SPL*> of the final network configuration for baseline results obtained in the low behavioral diversity parameter region ($d \leq 0.1$, $r_s \leq 0.1$, $r_w \leq 0.1$). Each dot represents a result of one simulation run. A black cubic trend curve is drawn to illustrate the social fragmentation transition between (<*CD*>, <*SPL*>) = (high, high) and (<*CD*>, <*SPL*>) = (low, low) behaviors. Meanwhile, the dashed circle indicates the missing (<*CD*>, <*SPL*>) = (high, low) behavior that does not occur when social constituents are behaviorally homogeneous.

Figure 2 shows the same (<*CD*>, <*SPL*>) plots for different standard deviations of $d$ (top), $r_s$ (bottom left) and $r_w$ (bottom right). It is seen that greater diversity of $d$ and $r_s$ helps maintain <*CD*> at higher levels. In addition, greater diversity of $d$ also helps lower <*SPL*> more (Fig. 2 top; orange/red dots), which corresponds to the (<*CD*>, <*SPL*>) = (high, low) behavior that was not previously recognized in the literature. Meanwhile, the effect of diversity of $r_w$ is not as clearly seen in this visualization compared to the other two parameters. These results strongly imply that having a broad distribution of cultural tolerance levels within society can help promote the coexistence of cultural diversity and structural connectivity.

Figure 3 shows typical final network configurations for two experimental settings. In a situation where culturally heterogeneous constituents remain connected (Fig. 3 right), constituents with different levels of cultural tolerance typically occupy different positions in the network. For example, less tolerating constituents tend to form clusters of their own, acting as cultural memory, while more tolerating ones tend to act as a “glue” to connect such culturally distinct clusters, serving as bridges.

We also conducted linear regression analysis to regress each of the two outcome variables onto the three experimental parameters (i.e., standard deviations of the three behavioral attributes; denoted as $\sigma d$, $\sigma r_s$, and $\sigma r_w$ below) and their interactions. Results are given in Equations (1) and (2), and their ANOVA tables are shown in Tables 1 and 2:

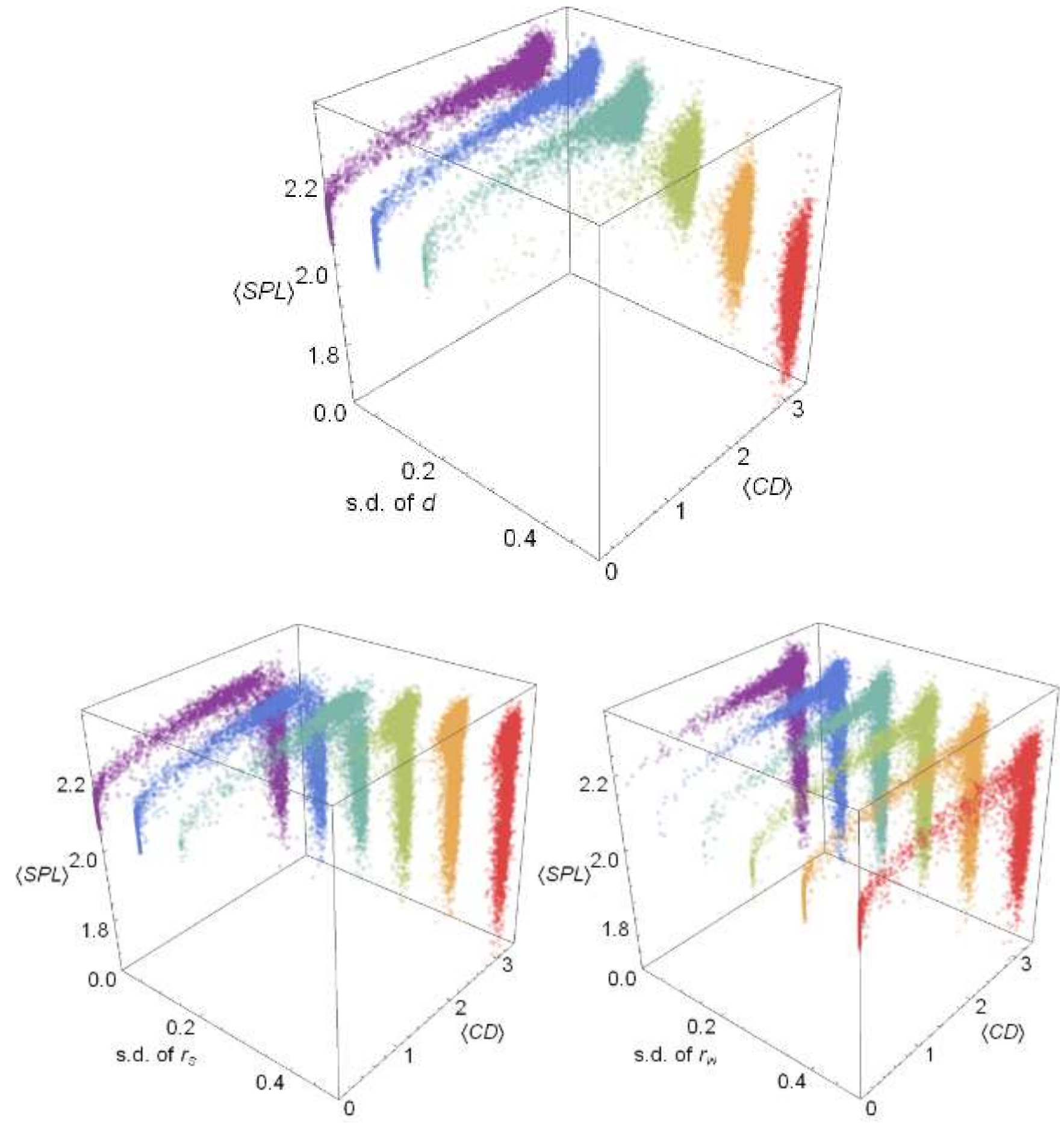

**Fig. 2.** 3D scatter plots showing the effect of each experimental parameter, i.e., standard deviation of *d* (top), $r_s$ (bottom left) and $r_w$ (bottom right), on two outcome measures <*CD*> and <*SPL*> of the final network configuration. Each dot represents a result of one simulation run, colored according to the parameter value.

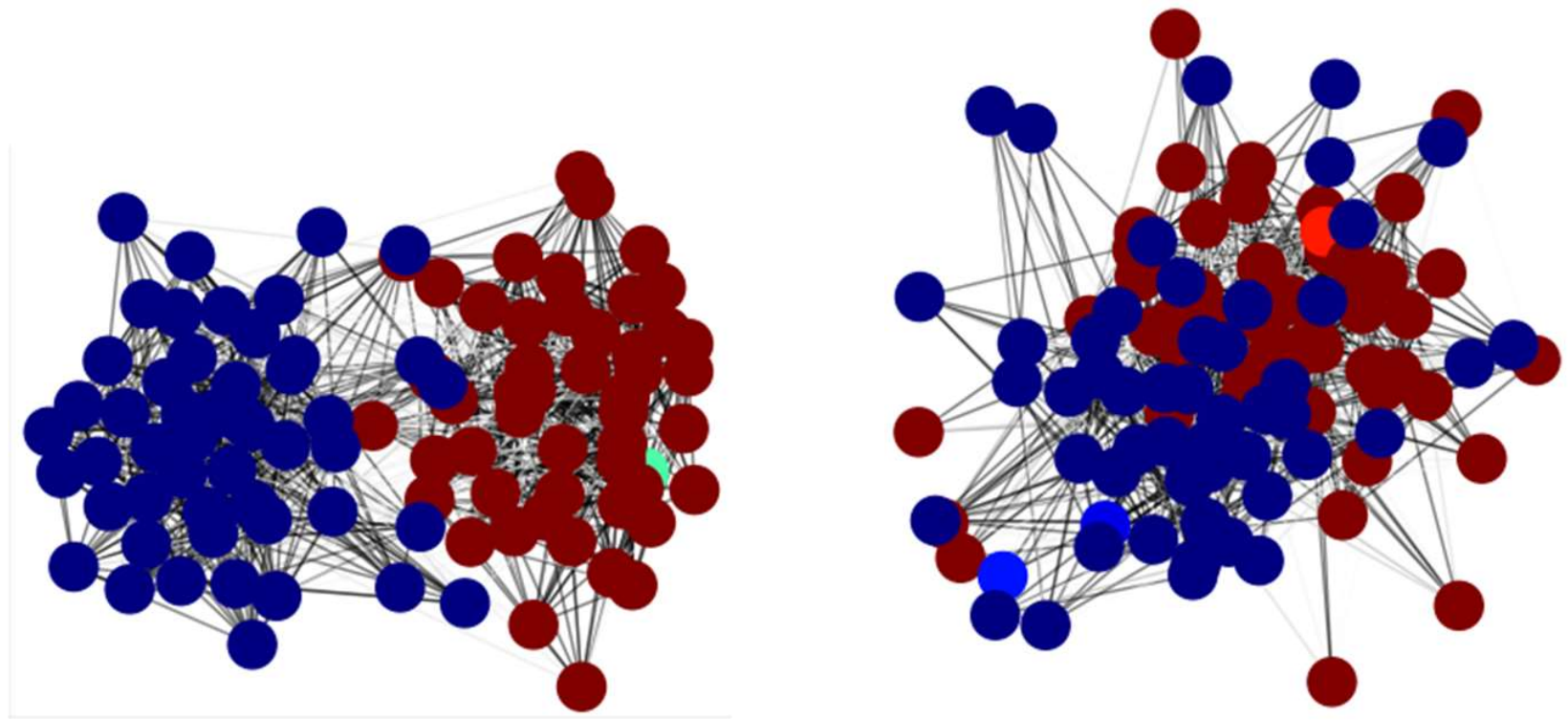

**Fig. 3.** Two examples of final network configuration. Left: Standard deviations of ($d$, $r_s$, $r_w$) = (0.0, 0.2, 0.2). Right: Standard deviations of ($d$, $r_s$, $r_w$) = (0.5, 0.2, 0.2). Node colors represent the individuals' cultural values (multidimensional vectors projected to a linear color scale).

$$<CD> \sim 1.87262 + 3.01908\ \sigma d + 2.97431\ \sigma r_s - 2.72074\ \sigma r_w - 8.95723\ \sigma d\ \sigma r_s + 3.60938\ \sigma d\ \sigma r_w + 3.90939\ \sigma r_s\ \sigma r_w \quad (5)$$

$$<SPL> \sim 2.31216 - 0.624629\ \sigma d + 0.0989771\ \sigma r_s - 0.178676\ \sigma r_w - 0.390949\ \sigma d\ \sigma r_s + 0.35265\ \sigma d\ \sigma r_w + 0.122775\ \sigma r_s\ \sigma r_w \quad (6)$$

**Table 1.** ANOVA table of linear regression of <*CD*> shown in Eq. (5). All terms were statistically extremely significant.

| | Sum of Squares | df | Mean Square | F | Sig. |
|---|---|---|---|---|---|
| $\sigma d$ | 1782.60 | 1 | 1782.60 | 9797.94 | p<.0001 |
| $\sigma r_s$ | 1847.26 | 1 | 1847.26 | 10153.30 | p<.0001 |
| $\sigma r_w$ | 445.64 | 1 | 445.64 | 2449.42 | p<.0001 |
| $\sigma d\ \sigma r_s$ | 1474.26 | 1 | 1474.26 | 8103.16 | p<.0001 |
| $\sigma d\ \sigma r_w$ | 239.38 | 1 | 239.38 | 1315.75 | p<.0001 |
| $\sigma r_s\ \sigma r_w$ | 280.83 | 1 | 280.83 | 1543.57 | p<.0001 |
| Error | 3928.56 | 21593 | 0.18 | | |
| Total | 9998.54 | 21599 | | | |

The linear terms in Eq. (5) imply that the average cultural distance (<*CD*>) is maintained by having the diversities of $d$ and $r_s$, while the diversity of $r_w$ has a negative effect on the cultural distance. Meanwhile, the linear terms in Eq. (6) indicate that the average shortest path length (<*SPL*>) is reduced by having the diversities of $d$ and $r_w$, while the diversity of $r_s$ has only a marginal (positive) effect on the average shortest path length.

**Table 2.** ANOVA table of linear regression of <*SPL*> shown in Eq. (6). All terms were statistically extremely significant. There were several simulation runs in which some individual nodes became disconnected, and such cases were excluded from the calculation (which is the reason that the dfs of Error and Total have smaller values than in Table 1).

| | Sum of Squares | df | Mean Square | F | Sig. |
|---|---|---|---|---|---|
| $\sigma d$ | 253.02 | 1 | 253.02 | 58391.20 | p<.0001 |
| $\sigma r_s$ | 0.64 | 1 | 0.64 | 148.78 | p<.0001 |
| $\sigma r_w$ | 2.26 | 1 | 2.26 | 521.21 | p<.0001 |
| $\sigma d\ \sigma r_s$ | 2.80 | 1 | 2.80 | 646.91 | p<.0001 |
| $\sigma d\ \sigma r_w$ | 2.28 | 1 | 2.28 | 526.55 | p<.0001 |
| $\sigma r_s\ \sigma r_w$ | 0.28 | 1 | 0.28 | 63.83 | p<.0001 |
| Error | 93.47 | 21571 | 0.00 | | |
| Total | 354.75 | 21577 | | | |

The nonlinear interaction terms in Eqs. (5) and (6) imply that the interaction between the diversities of $d$ and $r_s$ has a negative effect on both outcome measures, while other interaction terms generally have positive effects on them. Their interactions were visualized in more detail in the 3D scatter/surface plots shown in Fig. 4. These plots illustrate that the interactions of diversity parameters are much more significant on <*CD*> than on <*SPL*>, and that greater diversity of either $d$ or $r_s$ maintain <*CD*> consistently at a higher level.

Among all the terms included in these regression models, the only term whose coefficients point to the (<*CD*>, <*SPL*>) = (high, low) direction is the diversity of cultural tolerance ($\sigma d$). This result suggests that enhancing the diversity of cultural tolerance has the single most effective way to achieve the social state that maintain high cultural diversity and high structural connectivity simultaneously.

## 5 Conclusions

In this brief paper, we computationally studied the effects of behavioral diversities of social constituents on the resulting cultural diversity and social connectivity using a networked agent-based simulation model. Our results indicated that allowing cultural tolerance levels to differ broadly within society helps promote the coexistence of cultural diversity and structural connectivity, which is a novel macroscopic state of the adaptive social network models that was previously not known in the literature.

Our key finding above is interesting and relevant to network science, complex systems and social/organizational sciences in a couple of distinct ways.

First, it offers a clear demonstration of the risk in assuming that agents in a social system are identical and homogeneous. In view of the complexity of real-world systems, such a simplification is apparently wrong, but it is still widely used in many complex systems/network models of social dynamics. Our results show that inclusion of variations in individual attributes even in the simplest manner may already have huge impacts on the macroscopic outcomes of the system's evolution.

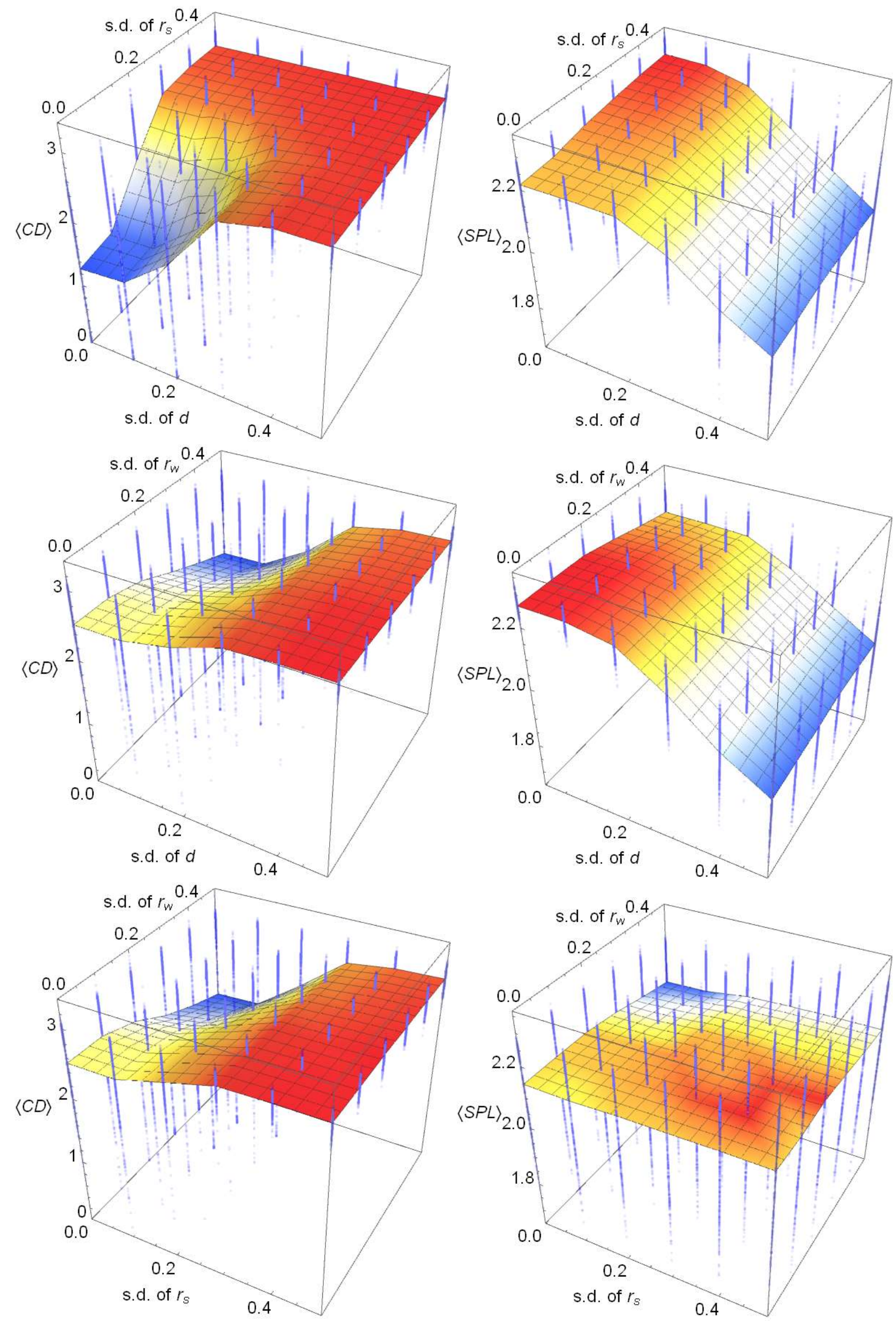

**Fig. 4.** 3D scatter plots showing the interactions between diversities of $d$ and $r_s$ (top), $d$ and $r_w$ (middle), and $r_s$ and $r_w$ (bottom), on outcome measures <*CD*> (left) and <*SPL*> (right). Each blue dot represents a result of one simulation run. Average trends are also shown as a surface.

Second, our results point out the importance of *behavioral* diversity, not demographic or other surface-level diversities that are often discussed in the context of social, organizational and political studies. In contrast to demographic properties that cannot be altered easily, behaviors of people are by a large part acquired traits, and therefore, they can be trained and modified through proper intervention. This indicates that our finding may eventually lead to some education/intervention strategies to promote the maintenance of informational diversity and communication, possibly enhancing the creativity and innovation of our society as a whole.

Third, we note that our finding indicates the importance of the *diversity* of cultural tolerance levels, and not the tolerance itself. In today's socio-political climate, cultural tolerance is highly encouraged, but our model does not imply that simply increasing the cultural tolerance levels globally within the social network would lead to beneficial outcomes. Additional experiments with globally enhanced cultural tolerance levels of all the constituents (i.e., simply increasing the value of $d$ for all individuals; results not shown here) did not generate the same outcome as presented in this paper, because such a condition would quickly lead to a loss of cultural diversity. This implies that, at least from the perspective of enhancing both informational diversity and communication, telling people to be just tolerating does not produce the desired outcomes. These findings and implications collectively illustrate the highly non-trivial nature of the cultural dynamics in our society.

This study is still far from completion, and there are several future tasks to conduct. One obvious limitation of our present model is that it is fairly complicated and is not suitable for mathematical analysis. We plan to develop a much more simplified model of the same adaptive social network dynamics so that its behavior can be analyzed and explained mathematically. The other important direction of future research is to compare the dynamics of these adaptive social network models with real-world data of information exchange in order to validate and revise the model assumptions. For this purpose, we are currently working on collecting empirical data of cultural dynamics from social media and other online/offline sources. Recent machine learning tools for content analysis [10,11] allow researchers to quantify similarities and differences between contents posted by users, and this information can be examined with regard to its potential correlation with temporal changes of future contents posted by the same users as well as their social relationships [12]. Our future goal is to use such empirical data to determine the role and importance of individual behavioral heterogeneity in real-world adaptive social networks, as predicted in the theoretical model presented in this paper.

## Acknowledgements

This work was supported by JSPS KAKENHI Grant Number 19H04220.